\documentclass{rspublic}

\usepackage{epsfig}

\def\half{\mbox{\small $\frac{1}{2}$}}

\begin{document}

   \title[QCD and Hadron Dynamics]{QCD and Hadron Dynamics}
   \author[Yu.L. Dokshitzer]{Yuri Dokshitzer}
   \affiliation{LPT, B\^at.\ 210, Universit\'e Paris Sud, 91405 Orsay,
France \\
and \\
PNPI, 188350, Gatchina, St.~Petersburg, Russia}
   \label{firstpage}
   \maketitle

\begin{flushright}
     LPT-Orsay-00/89\\
     August 2000 
\end{flushright}

   \begin{abstract}{perturbative QCD, multiple hadroproduction, colour
   confinement} Perturbative QCD predicts and describes various
   features of multihadron production. An amazing similarity between
   observable hadron systems and calculable underlying parton
   ensembles justifies the attempts to use the language of quarks and
   gluons down to small momentum scales, to approach the profound
   problems that are commonly viewed as being entirely
   non-perturbative.  \end{abstract}

\section{Introduction}

Quantum Chromodynamics is a wonder.  Take free quarks, supply them
with colour degrees of freedom, demand invariance with respect to the
``repainting'' quark fields arbitrarily in each point in space-time
--- and you get the unique QCD Lagrangian describing interacting
quarks and gluons.  Having done that, you (are supposed to) have the
whole hadron world in your hands!  Such a beauty and ambition is hard
to match.

At the same time, it is worth remembering that QCD is
probably the strangest of theories in the history of modern physics.
On one hand, the striking successes of QCD-based phenomenology leave
no doubt that QCD is indeed the microscopic theory of hadrons
and their interactions.
On the other hand, the depth of the conceptual problems that one faces
in trying to formulate QCD as a respectable Quantum Field Theory is
unprecedented.

QCD nowadays has a split personality. 
It embodies ``hard'' and ``soft'' physics, both being hard subjects, 
and the softer the harder. 

Until recently QCD studies were concentrated on small-distance
phenomena, observables and characteristics that are as insensitive to
large-distance confinement physics as possible.  This is the realm of
``hard processes'' in which a large momentum transfer $Q^2$, either
time-like $Q^2\gg 1$~GeV$^2$, or space-like $Q^2\ll -1$~GeV$^2$, is
applied to hadrons in order to probe their small-distance quark-gluon
structure.

High-energy annihilation $e^+e^-\to$~hadrons, deep inelastic
lepton-hadron scattering (DIS), production in hadron-hadron collisions
of massive lepton pairs, heavy quarks and their bound states, large
transverse momentum jets and photons are classical examples of hard
processes.

%    HERA   inclusion
%

Perturbative QCD (PT QCD) controls the relevant cross sections and, to
a lesser extent, the structure of final states produced in hard
interactions.  Whatever the hardness of the process, it is hadrons,
not quarks and gluons, that hit the detectors.  For this reason alone,
the applicability of the PT QCD approach, even to hard processes, is far
from being obvious.  One has to rely on plausible arguments
(completeness, duality) 
and look for observables that are less vulnerable towards our
ignorance about confinement. 

In particle physics a discovery of 
a class of  
%
%an important realization of the fact that some 
animals that are {\em more equal}\/ 
than the others is due to Sterman \& Weinberg (1977).
They introduced an important notion of 
Collinear-and-Infrared Safety. 

An observable is granted the CIS status if it can be calculated in
terms of quarks and gluons treated as real particles (partons),
without encountering either collinear ($\theta\to0$) or infrared
($k_0\to 0$) divergences.  The former divergence is a standard feature
of (massless) QFT with dimensionless coupling, the latter is typical
for massless vector bosons (photons, gluons).

This classification is more than mere zoology.  Given CIS quantity,
we expect its PT QCD value {\em predictable}\/ in the quark-gluon
framework to be directly comparable with its {\em measurable}\/ value
in the hadronic world.  For this reason the CIS observables are
the preferred pets of QCD practitioners. 

To give an example, we cannot deduce from the first principles parton
distributions inside hadrons (PDF, or structure functions). 
However, the rate of their $\ln Q^2$-dependence (scaling violation)
is an example of a CIS measure and stays under PT QCD jurisdiction. 

Speaking about the final state structure, we cannot predict, say, the
kaon multiplicity or the pion energy spectrum.  However, one can
decide to be not too picky and concentrate on global characteristics
of the final states rather than on the yield of specific hadrons.
Being sufficiently inclusive with respect to final hadron species, one
can rely on a picture of the energy-momentum flow in hard collisions
supplied by PT QCD --- the jet pattern.

There are well elaborated procedures for counting jets (CIS jet
finding algorithms) and for quantifying the internal structure of jets
(CIS jet shape variables). They allow the study of the gross features
of the final states while staying away from the physics of
hadronisation. Along these lines one visualizes asymptotic freedom,
checks out gluon spin and colour, predicts and verifies scaling
violation pattern in hard cross sections, etc.\ 
These and similar checks have constituted the basic QCD tests of the
past two decades.

This epoch is over. 
Now the High Energy Particle physics community is trying to probe
genuine confinement effects in hard processes to learn more about
strong interactions.  
The programme is ambitious and provocative.
Friendly phenomenology keeps it afloat and feeds our hopes of
extracting valuable information about physics of hadronisation.
%\cite{EPS}.

% HERA inclusion
%
%
The r\^ole of HERA in this quest is difficult to overestimate.  She
has already taught us a lot about the structure of the proton, its
quark and gluon content, and brought back into limelight, in quite a
dramatic fashion, such basic issues as the Pomeron, diffraction,
unitarity.

It is true that $e^+e^-$ annihilation provides the cleanest
environment for studying the QCD interactions and hadron jets.
The advantage of HERA, however, is that in the DIS environment the
total energy of the collision and the hardness of the interaction are
not strictly
%kinematically
linked as is the case for the point-like $e^+e^-$ annihilation. 

The power of HERA, the DIS goddess, lies in her ability to study
various scales, from very large $Q^2$ down to moderate and small
momentum transfers, thus probing an interface between hard and soft
physics.  While it can be said that $e^+e^-$ will always remain the
best ground for {\em testing}\/ QCD, DIS is better suited for {\em
understanding}\/ it.

\section{Multihadron production and QCD}

In general, there are three ways to probe the small-distance
hadron structure.  
\begin{itemize}
\item[vac$\to$hadrons:] High energy vacuum excitation producing
hadrons, like in (but not exclusively) $e^+e^-$ annihilation.   
\item[vac+h$\to$hadrons:] Large momentum transfer excitation of an
initial hadron by a sterile 
%(``vacuum'') 
probe, like in deep inelastic lepton-hadron scattering (DIS).
\item[h+h$\to$hadrons:] Production of large--$p_\perp$ 
hadrons in hadron-hadron collisions. (Here sterile probes can be
employed in the final state as well, e.g.\ massive
lepton pairs and/or large--$p_\perp$ photons.)
\end{itemize}
Copious production of hadrons is typical for all these processes.  On
the other hand, at the microscopic level, multiple quark-gluon
``production'' is to be expected as a result of QCD bremsstrahlung ---
gluon radiation accompanying abrupt creation/scattering of colour  
partons. 

Is there a correspondence between observable hadron and calculable
quark-gluon production?

An indirect evidence 
%The fact 
that gluons are there, and that they behave, can be obtained from the study
of the scaling violation pattern.  QCD quarks (and gluons) are not
point-like particles, as the orthodox parton model once assumed. Each
of them is surrounded by a proper field coat --- a coherent virtual
cloud consisting of gluons and ``sea'' $q\bar{q}$ pairs.  A hard probe
applied to such a dressed parton breaks coherence of the
cloud. Constituents of these field fluctuations are then released as
particles accompanying the hard interaction. The harder
the hit, the larger an intensity of bremsstrahlung and, therefore, the
fraction of the energy-momentum of the dressed parton that the 
bremsstrahlung quanta typically carry away. 
Thus we should expect, in particular, that the probability that a
``bare'' core quark carries a large fraction 
%$x\sim1$ 
of the energy of its dressed parent will decrease with increase of
$Q^2$.  And so it does.

The logarithmic scaling violation pattern in DIS structure
functions is well established and meticulously follows the QCD
prediction based on the parton evolution picture.

\begin{figure}[ht]\label{delphiscaling}
\epsfig{file=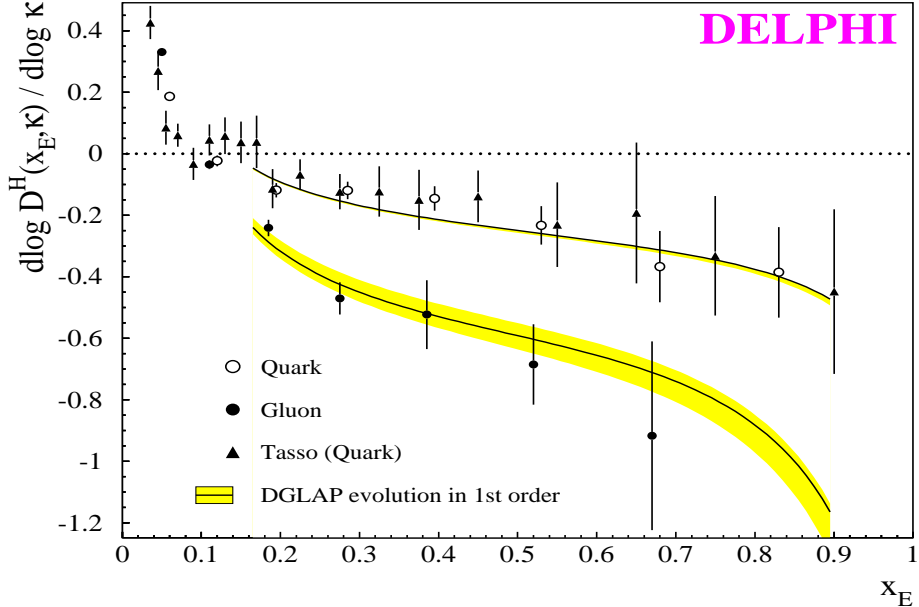,height=8cm,width=12cm}
   \caption{Scaling violation rates in inclusive hadron distributions
   from gluon and quark jets (Hamacher {\em et. al} 1999).
% and their ratio compared with that of the
%   ``colour charges'' $C_A/C_F=9/4$.
}
\end{figure}

In DIS we look for a ``bare'' quark inside a target dressed one. 
In $e^+e^-$ hadron annihilation at large energy $s=Q^2$ the chain of
events is 
reversed.
%opposite.  
Here we produce instead a bare quark with energy $Q/2$, which then
``dresses up''.  In the process of restoring its proper field-coat our
parton produces (a controllable amount of) bremsstrahlung radiation
which leads to formation of a hadron jet.  Having done so, in the end
of the day it becomes a constituent of one of the hadrons that hit the
detector.  Typically, this is the leading hadron. However, the
fraction $x_E$ of the initial energy $Q/2$ that is left to the leader
depends on the amount of accompanying radiation and, therefore, on
$Q^2$ (the larger, the smaller). In fact, the same rule (and the same
formula) applies to the scaling violation pattern in $e^+e^-$
fragmentation functions (time-like parton evolution) as to that in the
DIS parton distributions (space-like evolution).

What makes the annihilation channel particularly interesting, is that
the present day experiments are so sophisticated that they provide us
with a near-to-perfect separation between quark- and gluon-initiated
jets (the latter being extracted from heavy-quark-tagged three-jet events).

In Fig.~1
%\ref{delphiscaling} 
a comparison is shown of the scaling violation rates in the hadron
spectra from gluon and quark jets, as a function of the hardness scale
$\kappa$ that characterizes a given jet (Hamacher {\em et al.} 1999).
For large values of $x_E\sim1$ the ratio of the logarithmic
derivatives is predicted to be close to that of the gluon and quark
``colour charges'', $C_A/C_F=9/4$. Experimentally, the ratio is
measured to be
\begin{equation}
  \label{eq:cacf}
\frac{C_A}{C_F}= 2.23  \pm 0.09_{\mbox{\scriptsize stat.}} \pm 
0.06_{\mbox{\scriptsize syst.}}.    
\end{equation}

\subsection{Mean parton and hadron multiplicities}

Since accompanying QCD radiation seems to be there, we can make a step
forward by asking for a {\em direct}\/ evidence: what is the fate of
those gluons and sea quark pairs produced via multiple initial gluon
bremsstrahlung followed by parton multiplication cascades?
Let us look at the $Q$-dependence of the mean hadron multiplicity, the
quantity dominated by relatively soft particles with $x_E\ll1$. 
This is the kinematical region populated by accompanying QCD radiation.

\begin{figure}[th]\label{delphimult}
\epsfig{file=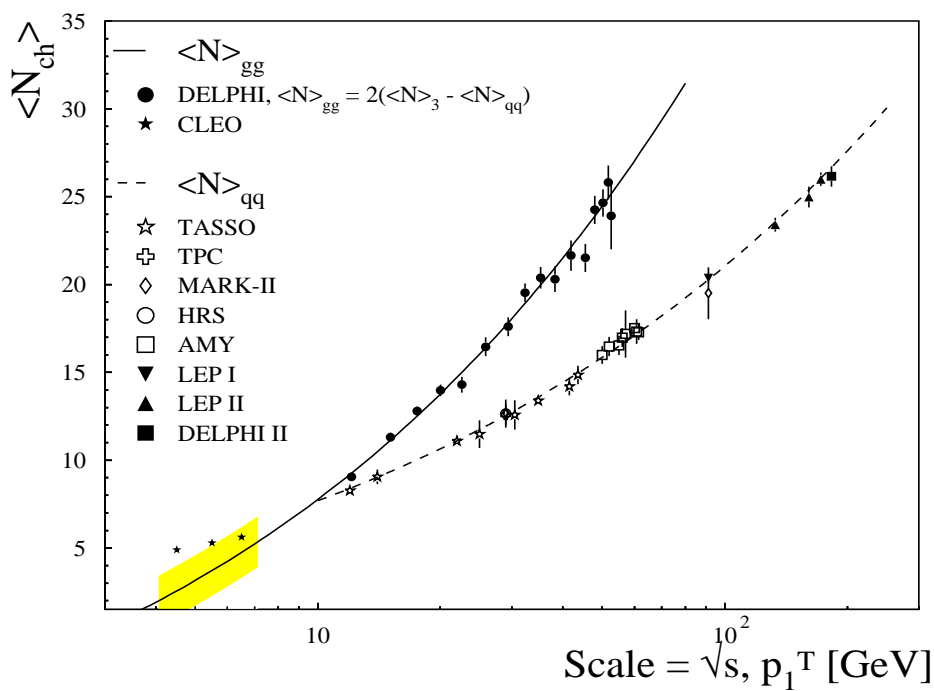,height=9cm,width=12cm}
   \caption{Charged hadron multiplicities in gluon and quark jets
%, and the ratio of slopes compared with that of the ``colour charges''
%$C_A/C_F=9/4$
 (DELPHI 1999).}
\end{figure}

Fig.~2
%\ref{delphimult} 
demonstrates that the hadron multiplicity increases with the hardness
of the jet proportional to
%that of 
the multiplicity of secondary gluons and sea quarks.  The ratio of the
slopes, once again, provides an independent measure of the ratio of
the colour charges, which is consistent with \eqref{eq:cacf} (DELPHI 1999):
\begin{equation}
\frac{C_A}{C_F}= 2.246  \pm 0.062_{\mbox{\scriptsize stat.}} \pm 
0.008_{\mbox{\scriptsize syst.}} \pm 0.095_{\mbox{\scriptsize theo.}}.      
\end{equation}

\subsection{Inclusive hadron distribution in jets}

Since the total numbers match, it is time to ask a more delicate
question about energy-momentum distribution of final hadrons versus
that of the underlying parton ensemble.  One should not be too picky
in addressing such a question.  It is clear that hadron-hadron
correlations, for example, will show resonant structures about which
the quark-gluon speaking PT QCD can say little, if anything, at the
present state of the art.  Inclusive single-particle distributions,
however, have a better chance to be closely related.
Triggering a single hadron in the detector, and a single parton on
paper, one may compare the structure of the two distributions
to learn about dynamics of hadronisation. 
 
% apply a duality consideration to argue that 

Inclusive energy spectrum of soft bremsstrahlung partons in QCD
jets has been derived in 1984 in the so-called MLLA --- the Modified Leading
Logarithmic Approximation (Dokshitzer \& Troyan 1984). 
This approximation takes into account all essential ingredients of
parton multiplication in the next-to-leading order. 
They are: parton splitting functions responsible for the energy balance in
parton splitting, the running coupling $\alpha_s(k_\perp^2)$
depending on the relative transverse momentum of the two offspring 
and exact angular ordering. 
The latter is a consequence of soft gluon coherence and plays an
essential r\^ole in parton dynamics.  In particular, gluon coherence
suppresses multiple production of very small momentum gluons. It is
particles with intermediate energies that multiply most efficiently.
As a result, the energy spectrum of relatively soft secondary partons
in jets acquires a characteristic hump-backed shape. 
The position of the maximum in the logarithmic variable 
$\xi=-\ln x$, the width of the hump and its height increase  
with $Q^2$ in a predictable way. 

The shape of the inclusive spectrum of all charged hadrons (dominated
by $\pi^{\pm}$) exhibits the same features. This comparison, pioneered
by Glen Cowan (ALEPH) and the OPAL collaboration,
% in mid 80's, ?????????????
has later become a standard test of analytic QCD predictions.
First scrutinized at LEP, the similarity of parton and hadron energy
distributions has been verified at SLC and KEK $e^+e^-$ machines, as
well as at HERA and Tevatron where hadron jets originate not from bare
quarks dug up from the vacuum by a highly virtual photon/$Z^0$ but
from hard partons kicked out from initial hadron(s).

In Fig.~3 (DELPHI) 
%
%   ?????????????
%
the comparison is made of the all-charged hadron spectra at various
annihilation energies $Q$ with the so-called ``distorted Gaussian''
fit (Fong \& Webber 1989) which employs the first four moments (the
mean, width, skewness and kurtosis) of the MLLA distribution around
its maximum.

\begin{figure}[h]\label{delspec}
\epsfig{file=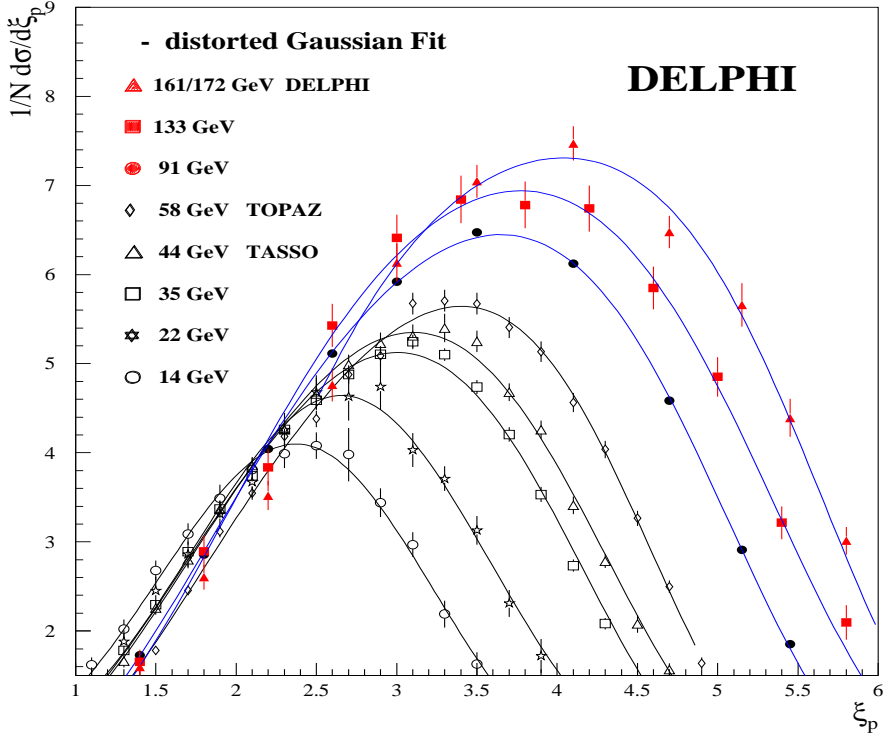,height=10cm,width=12cm}
   \caption{Inclusive energy distribution of charged hadrons in
jets produced in $e^+e^-$ annihilation   }
\end{figure}

Shall we say, a (routine, interesting, wonderful) check of yet another
QCD prediction?  Better not. Such a close similarity offers a deep
puzzle, even a worry, rather than a successful test.  Indeed, after a
little exercise in translating the values of the logarithmic variable
$\xi=\ln(E_{\mbox{\scriptsize jet}}/p)$ in Fig.~3 into GeVs you will
see that the actual hadron momenta at the maxima are, for example,
$p$=$\half Q\cdot e^{-\xi_{\max}}\simeq$~0.42, 0.85 and 1.0~GeV for
$Q$=14, 35~GeV and at LEP-1, $Q$=91~GeV.  Is it not surprising that
the PT QCD spectrum is mirrored by that of the pions (which
constitute 90\%\ of all charged hadrons produced in jets) with momenta
well below 1~GeV?!

For this very reason the observation of the parton-hadron similarity
was initially met with a serious and well grounded scepticism: it
looked more natural (and was more comfortable) to blame the finite
hadron mass effects for falloff of the spectrum at large $\xi$ (small
momenta) rather than seriously believe in applicability of the
PT QCD consideration down to such disturbingly small
momentum scales.

This worry has been recently answered. 
Andrey Korytov (CDF)
%(19??) 
was the first to hear a theoretical hint 
%(Dokshitzer et al. 19??) 
and carry out a study of the energy distribution of hadrons produced
inside a restricted angular cone $\Theta$ around the jet axis.
Theoretically, it is not the energy of the jet but the maximal parton
transverse momentum inside it, $k_{\perp\max}\simeq
E_{\mbox{\scriptsize jet}}\sin\frac{\Theta}{2}$, that determines the
hardness scale and thus the yield and the distribution of the
accompanying radiation.

This means that by choosing a small opening angle one can study
relatively small hardness scales but in a cleaner environment: due to
the Lorentz boost effect, eventually all particles that form a short
small-$Q^2$ QCD ``hump'' are now relativistic and concentrated at the
tip of the jet.

For example, selecting hadrons inside a cone $\Theta\simeq 0.14$
around an energetic quark jet with $E_{\mbox{\scriptsize jet}}\simeq
100$~GeV (LEP-2) one should see that very ``dubious'' $Q=14$~GeV curve
in Fig.~3 but now with the maximum boosted from 450~MeV into a
comfortable 6 GeV range.

In the CDF Fig.~4 (A. Korytov 1996, personal communication; Goulianos
1997, see also Safonov 1999) a close similarity between the hadron
yield and the full MLLA parton spectra
%correspondence 
can no longer be considered accidental and be attributed to
non-relativistic kinematical effects.

\begin{figure}[ht]\label{cdfspec}
\epsfig{file=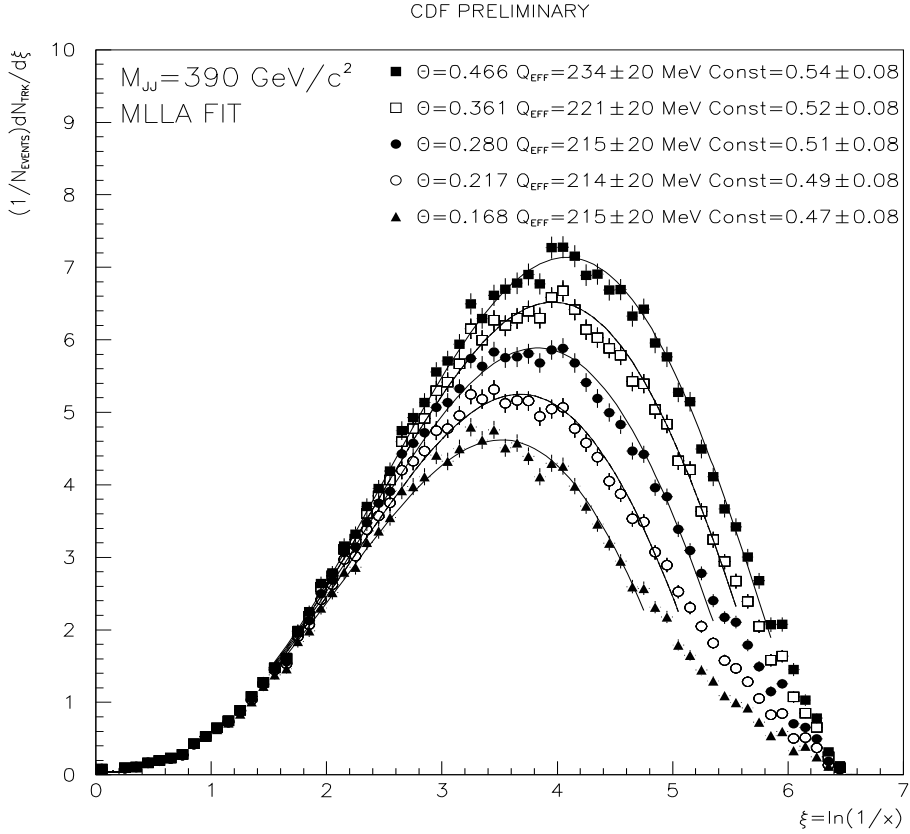,height=11cm,width=12cm}
   \caption{Inclusive energy distribution of charged hadrons in 
large--$p_\perp$ jets  
(Goulianos 1997).
}
\end{figure}

\subsection{Brave gluon counting}

Modulo $\Lambda_{\mbox{\scriptsize QCD}}$, there is only one unknown 
in this comparison, namely, the overall normalisation of the 
spectrum of hadrons relative to that of partons (bremsstrahlung gluons).  

Strictly speaking, there should/could have been another free
parameter, the one which quantifies one's bravery in applying the PT
QCD dynamics. It is the minimal transverse momentum cutoff in parton
cascades, $k_\perp>Q_0$.  The strength of successive $1\to2$ parton
splittings is proportional to $\alpha_s(k_\perp^2)$ and grows with
$k_\perp$ decreasing.  The necessity to terminate the process at some
low transverse momentum scale where the PT coupling becomes large (and
eventually hits the formal ``Landau pole'' at
$k_\perp=\Lambda_{\mbox{\scriptsize QCD}}$) seems imminent.
Surprisingly enough, it is not.

Believe it or not, the inclusive parton energy distribution turns out
to be a CIS QCD prediction.  Its crazy $Q_0=\Lambda_{\mbox{\scriptsize
    QCD}}$ limit (the so-called ``limiting spectrum'') is shown by
solid curves in Fig.~4.

Choosing the minimal value for the collinear parton cutoff $Q_0$ can
be looked upon as shifting, as far as possible, responsibility
for particle multiplication in jets to the PT dynamics.
This brave choice can be said to be dictated by experiment, in a
certain sense. Indeed, with increase of $Q_0$ the parton parton
distributions {\em stiffen}\/ (parton energies are limited from below
by the kinematical inequality $xE_{\mbox{\scriptsize jet}}\equiv k \ge
k_\perp > Q_0$).  The maxima would move to larger $x$ (smaller
$\xi$), departing from the data.

\begin{figure}[ht]\label{cdfmax}
\epsfig{file=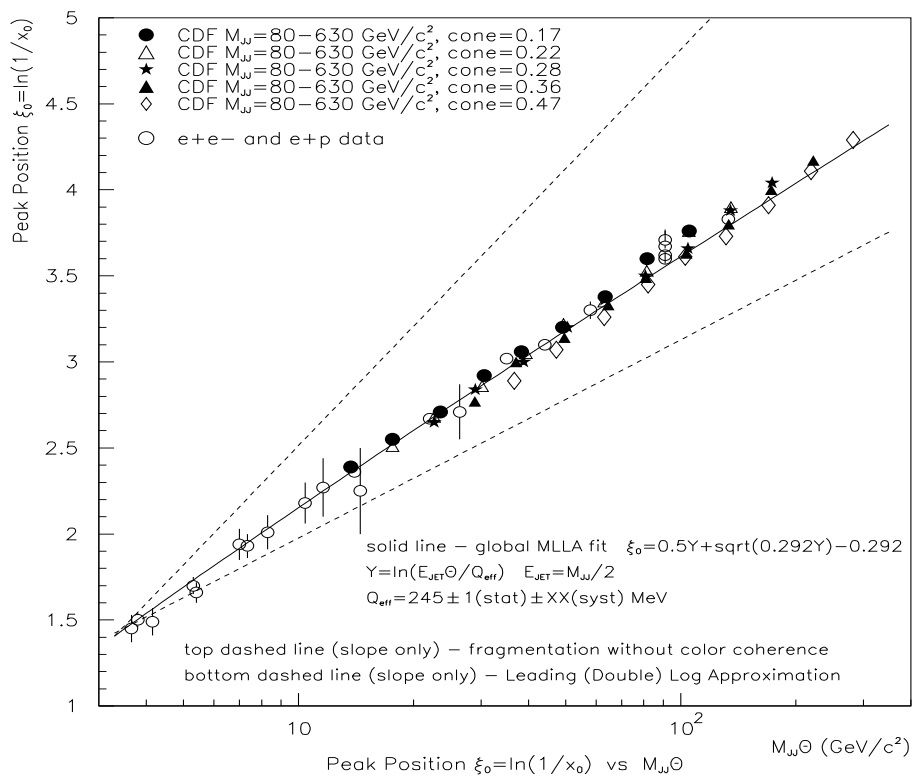,height=10.5cm,width=12cm}
   \caption{The position of the maximum versus the analytic MLLA
     prediction (Safonov 1999).}
\end{figure}

A clean test of ``brave gluon counting'' is provided by
Fig.~5 where the position of the hump, which is insensitive to the
overall normalisation, 
%as seen in $e^+e^-$, DIS and $pp$  
is compared with the parameter-free MLLA QCD prediction
(Safonov 1999).

A formal explanation of the tolerance of the {\em shape}\/ of inclusive
parton spectra to the dangerous small-$k_\perp$ domain can be found in
the proceedings of the last year Blois conference (Dokshitzer 1999).

To put a long story short, decreasing $Q_0$ we start to lose control
of the interaction intensity of a parton with a given $x$ and
$k_\perp\sim Q_0$ (and thus may err in the overall production rate).
However, such partons do not branch any further, do not produce any
soft offspring, so that the {\em shape}\/ of the resulting energy
distribution remains undamaged.  
Colour coherence plays here a crucial r\^ole.

It is important to realize that knowing the spectrum of {\em partons},
even knowing it to be a CIS quantity in certain sense, does not
guarantee on its own 
the
predictability of the {\em hadron}\/ spectrum.  It is easy to imagine
a world in which each quark and gluon with energy $k$ produced at the
small-distance stage of the process would have dragged behind its
personal ``string'' giving birth to $\ln k$ hadrons in the final state
(the Feynman plateau).
The hadron yield then would be given by a convolution of the parton
distribution with a logarithmic energy distribution of hadrons from
the parton fragmentation. 

If it were the case, each parton would have contributed to the yield
of non-relativistic hadrons and the hadron spectra would peak at much
smaller energies, $\xi_{\max}\simeq \ln Q$, in a spectacular
difference with experiment.

Physically, it could be possible if the non-perturbative (NP)
hadronisation physics did not respect the basic rule of the
perturbative dynamics, namely, that of colour coherence.

There is nothing wrong with the idea of convoluting time-like parton
production in jets with the inclusive NP parton$\to$hadron
fragmentation function, the procedure which is similar to convoluting
space-like parton cascades with the NP initial parton distributions
in a target proton to describe DIS structure functions.

What the nature is telling us, however, is that this NP fragmentation
has a finite multiplicity and is {\em local}\/ in the momentum space.
Similar to its PT counterpart, the NP dynamics has a short memory: the
NP conversion of partons into hadrons occurs locally in the
configuration space.

In spite of a known similarity between the space- and time-like parton
evolution pictures ($x\sim1$), there is an essential difference
between {\em small}--$x$ physics of DIS structure functions and the jet
fragmentation.
In the case of the space-like evolution, in the limit of small
Bjorken--$x$ the problem becomes essentially non-perturbative and PT
QCD loses control of the DIS cross sections (Mueller 1997, Camici \&
Ciafaloni 1997).  On the contrary, studying small Feynman--$x$
particles originating from the time-like evolution of jets offers a
gift and a puzzle: all the richness of the confinement dynamics
reduces to a mere overall normalisation constant.

The fact that even a legitimate finite smearing due to hadronisation
effects does not look mandatory makes one think of a deep duality
between the hadron and quark-gluon languages applied to such a global
characteristic of multihadron production as an inclusive energy
spectrum.

Put together, the ideas behind the brave gluon counting are known as
the hypothesis of Local Parton-Hadron Duality. Experimental evidence
in favour of LPHD is mounting, and so is list of challenging questions
to be answered by the future quantitative theory of colour
confinement.

\subsection{QCD Radiophysics}

Another class of multihadron production phenomena speaking in favour
of LPHD is the so-called inter-jet physics. It deals with  
particle flows in the angular regions between jets in various 
multi-jet configurations.
These particles do not belong to any particular jet, and their
production, at the PT QCD level, is governed by coherent soft gluon
radiation off the multi-jet system as a whole. Due to QCD coherence, 
these particle flows are insensitive to internal structure of 
underlying jets. The only thing that matters is the colour topology of 
the primary system of hard partons and their kinematics. 

The ratios of particle flows in different inter-jet valleys are given
by parameter-free PT QCD predictions and reveal the so-called
``string'' or ``drag'' effects. For a given kinematical jet
configuration such ratios depend only on the number of colours
($N_c$).

For example, the ratio of the multiplicity flow between a quark
(antiquark) and a gluon to that in the $q\bar{q}$ valley in symmetric
(``Mercedes'') three-jet $q\bar{q}g$ $e^+e^-$ annihilation events is
predicted to be
\begin{equation}
  \label{eq:qqg}
  \frac{dN^{(q\bar{q}g)}_{qg}}{dN^{(q\bar{q}g)}_{q\bar{q}}} 
  \simeq  \frac{5N_c^2-1}{2N_c^2-4}= \frac{22}{7}.
\end{equation}
Comparison of the denominator with the density of radiation in the
$q\bar{q}$ valley in $q\bar{q}\gamma$ events with a gluon jet replaced
by an energetic photon results in
\begin{equation}
  \label{eq:qqgam}
  \frac{dN^{(q\bar{q}\gamma)}_{q\bar{q}}}{dN^{(q\bar{q}g)}_{q\bar{q}}}  
  \simeq \frac{2(N_c^2-1)}{N_c^2-2}= \frac{16}{7}.
\end{equation}
%to be compared with the experimental ratio $2.3\pm 0.2$  OPAL ??? 
Emitting an energetic gluon off the initial quark pair depletes
accompanying radiation in the backward direction: colour is {\em
dragged}\/ out of the $q\bar{q}$ valley. This destructive interference
effect is so strong that the resulting multiplicity flow falls below
that in the least favourable direction transversal to the three-jet
event plane:
\begin{equation}
  \label{eq:qqtransv}
  \frac{dN^{(q\bar{q}\gamma)}_{\perp}} {dN^{(q\bar{q}g)}_{q\bar{q}}}  
  \simeq \frac{N_C+2C_F}{2(4C_F-N_c)}= \frac{17}{14}.
\end{equation}
At the level of the PT accompanying gluon radiation (QCD radiophysics)
such predictions are quite simple and straightforward to derive.  The
strange thing is, that these and many similar numbers are being seen
experimentally.  The inter-jet particles flows we are discussing are
dominated, at present energies, by pions with typical momenta in the
100--300 MeV range!  The fact that even such soft junk follows the PT
QCD rules is truly amazing.

\subsection{Soft confinement}

Honestly speaking, it makes little sense to treat few-hundred-MeV
gluons as PT quanta. What hadron energy spectra and string/drag
phenomena are trying to tell us is that the production of hadrons is
driven by the strength of the underlying colour fields generated by
the system of energetic partons produced in a hard interaction.
Pushing PT description down into the soft gluon domain is a mere tool
for quantifying the strengths of the colour field.

Mathematical similarity between the parton and hadron energy and
angular distributions means that confinement is very soft and gentle.
As far as the global characteristics of final states are concerned,
there is no sign of strong forces at the hadronisation stage which
forces 
would re-shuffle particle momenta when the transformation from
coloured quarks and gluons to blanched hadrons occurs.  (For a recent
review of MLLA-LPHD issues see Khoze \& Ochs 1997.)

This observation goes along with what we have learnt from studying DIS,
with special thanks to HERA which taught us that proton
is truly fragile. It suffices to kick it with 1~GeV momentum transfer,
or even less, and it is blown to pieces.  

It seems that what keeps a proton together is not any strong forces
between the quarks but merely quantum mechanics: the proton just
happened to be 
the
ground state with a given well conserved quantum number (baryon
charge). It is interesting to see how easy is it to break a proton. To
achieve that it is not even necessary to kick it hard. A soft scratch
(or rather two) is enough to do the job.

\begin{figure}[h]\label{stopping}
\epsfig{file=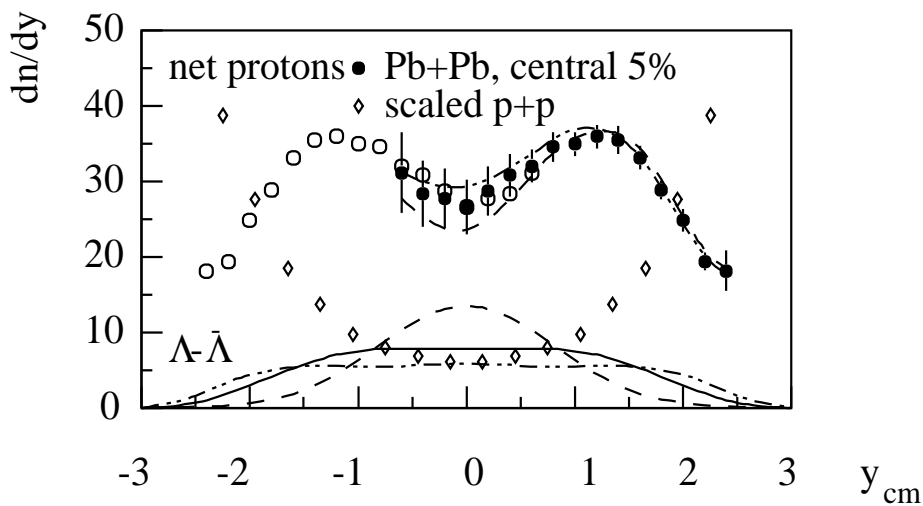,height=6.7cm,width=13cm}
   \caption{Proton ``stopping'' as seen at CERN by NA--49 (1999)}
\end{figure}

There is no sign of advocated fragility in a normal (minimum bias,
soft) high energy proton-proton scattering.  The famous leading
particle effect shows that a projectile protons stays intact in the
final state and carries away a major fraction of the incident momentum
(the net proton spectrum ``scaled $p+p$ '' in Fig.~6).  This is not
surprising. In a typical $pp$ interaction it is only one of the
valence quarks of the proton that scatters. Internal coherence of the
spectator quark pair remains undisturbed. In these circumstances the
proton splits into a triplet quark and a spectator diquark which is
in a colour anti-triplet state. On the hadronisation stage, the former
picks up an antiquark and turns into a meson carrying, roughly,
$z\simeq1/3$ of the initial proton momentum, while the diquark (colour
equivalent of a $\bar{q}$) picks up a quark forming a leading baryon
($z\simeq2/3$).

It suffices, however, to organize a {\em double}\/ scattering within a
life-time of the intrinsic proton fluctuation in order to destroy the
proton coherence completely (including that of the diquark which
remained intact after the first scratch).  Now the three
quark-splinters 
of the proton separate as independent triplet charges and normally
convert into three ``leading'' mesons (carrying $z\simeq1/3$ each) in
the final state. The proton decays, for example, into
$$ 
p(1) \to \pi^+(1/3) + \pi^-(1/3) + K^+(1/3) + \ldots 
$$
with the baryon quantum number sinking in the sea.
%, to reemerge as a small-momentum baryon  

This is what seems to be going on in the ion-ion scattering as shown
in Fig.~6. Disappearance of leading protons is known as ``stopping''
in the literature.  This I believe is an inadequate name: there is no
way to {\em stop}\/ an energetic particle, especially in soft
interaction(s).  Relativistic quantum field theory is more tolerant to
changing particle identity than to allowing a large transfer of
energy-momentum (recall relativistic Compton 
%scattering as an example,
where the {\em backward}\/ scattering dominates: electron turns into a
forward photon, and vice versa).

If this heretic explanation of the ``stopping'' as proton instability
is correct, the same phenomenon should be seen in the proton
hemisphere of proton-nucleon collisions and even in $pp$.  As we know,
here there are leading protons. However, this is true on {\em
average}.  Even in $pp$ collisions one can enforce multiple scattering
(and thus full proton breakup) by selecting rear events, e.g.\  with
larger than average final state multiplicity.  

In all these cases ($pp$, $pA$, $AB$) ``proton decay'' should be
accompanied by an enhanced strangeness production. Collecting
experimental evidence in favour of proton instability is underway
(Fischer 2000).

Soft hadronisation, likely absence of strong inter-parton forces,
fragile hadrons --- can it be reconciled with confinement in the
first place?
To the best of my knowledge, Gribov Super-Critical Light-Quark
Confinement theory (GSCC) is the only scenario to offer a natural
explanation to the puzzling phenomenology of multihadron production
discussed above. 

Light quarks are crucial for GSCC. If I was not so ignorant in
theology (and was not brought up to hate philosophy), we could spend
some time discussing, has it not been done on purpose that 
%the 
God supplied us with very light (practically massless $u$ and $d$)
quarks in order to make the hadron world easier to understand?

It is clear without going into much mathematics that the presence of
light quarks is sufficient 
for preventing 
the colour forces from growing real big: dragging away a heavy quark
we soon find ourselves holding a blanched $D$-meson
instead. The light quark vacuum is eager to screen any separating colour
charges.

The question becomes quantitative: how strong is strong? How much of a
tension 
does one need
to break the vacuum and organize such a screening?  In the early 90-s
Gribov has shown that in a theory with a Coulomb-like interaction
between light fermions it suffices to have the coupling exceeding the
critical value,
\begin{equation}
\label{crit}
  %(C_F) 
\frac{\alpha_{\mbox{\scriptsize crit.}}}{\pi} \simeq 1-\sqrt\frac23 \,,
\end{equation}
to have super-critical binding, restructuring of the PT vacuum, chiral
symmetry breaking and, likely, confinement (Gribov 1999 and references
therein).  The word {\em super-critical}\/ refers to the known QED
phenomenon of so-called super-critical atoms. Dirac energy levels of
an electron in a point-like Coulomb field of an ion with $Z> 137$
become complex. Classically, the electron ``falls into the
centre''. Quantum-mechanically, it also falls, but into the Dirac sea:
the ion becomes unstable and gets rid of an excessive electric charge
by emitting a positron (Pomeranchuk \& Smorodinsky 1945).

In the QCD context, with the colour factor $C_F=4/3$ applied to the
l.h.s.\ of \eqref{crit}, the critical coupling becomes
\begin{equation}
\label{QCDcrit}
\left(\frac{\alpha_{s}}{\pi}\right)_{\mbox{\scriptsize crit.}} \simeq 0.137\,.
\end{equation}
This number, apart from being easy to memorize, has another important
quality: it is numerically small. 
Gribov's ideas, 
%if understood and pursued, 
being understood and pursued,
offer an intriguing possibly to address all the diversity and
complexity of the hadron world from within the field theory with a
reasonably small effective interaction strength (read: {\em
perturbatively}\/).

\section{Probing NP dynamics with PT tools}

Can one talk about QCD coupling at small momentum scales? To answer
such a question positively is not easy. Apart from courage, one needs
to design some more or less definite prescription for quantifying an
interaction strength at large distances where the very objects that
are supposed to interact kind of don't exist!  The best collection of
arguments I could come up with, convincing or not, can be found in the
proceedings of the HEP Vancouver conference (Dokshitzer 1998).

In recent years first steps have been made towards a joint technology
for triggering and quantifying non-perturbative effects in CIS
observables, both in ``Euclid-translatable'' cross sections and in the
essentially Minkowskian characteristics of hadronic final states.
The fact that the CIS observables are calculable in PT QCD (that is,
remain finite when the collinear QCD cutoff $\mu$ is set to zero) does
not imply that they are completely insensitive to NP dynamics.  This
only means that the genuine NP effects in CIS quantities manifest
themselves as finite power suppressed corrections proportional to
$(\mu^2/Q^2)^p\log^q (Q^2/\mu^2)$ with $p>0$.

Simply by examining PT Feynman diagrams, one can find the exponents
$p,q$ for different observables. Knowing the leading power $p$ is
already useful: it tells us how (in)sensitive to confinement physics a
given observable is. 

More ambitious a programme aims at the {\em magnitudes}\/ of
power-suppressed contributions to hard cross sections and jet shape
variables. (For references and a history of insights, phenomenological
achievements and conceptual and numerical mistakes, this young subject
is so rich with, the reader is invited to look into the proceedings of
the Vancouver--1998 and Blois--1999 conferences.)

The magnitudes of the power-suppressed terms can be related with the
behaviour of the coupling $\alpha_s$ in the infrared.  Whatever the
definition, it is thought to be a {\em universal}\/ function that
characterizes, in an effective way, the strength of the QCD
interaction all the way down to small momentum scales. Given this
universality, it becomes possible to {\em predict}\/ the ratios of the
$Q^{-2p}$ contributions to observables belonging to the same class
$p$.

In particular, the characteristic NP parameter
\begin{equation} 
 \alpha_0 = \frac1{\mu_I} \int_0^{\mu_I} dk\, \alpha_s(k^2), \quad
 (\mu_I=2~\mbox{GeV})
\end{equation}
is conveniently used to quantify the NP hadronisation effects in CIS
jet shapes, many of which belong to the $p=\half$ class, i.e.\ exhibit
large $1/Q$ power corrections.
These include the thrust $T$, the so called $C$-parameter, 
invariant jet masses $M_J$ and $M_H$ (heavy-jet mass), 
the jet broadenings $B_T$ and $B_W$ (wide-jet broadening). 
\begin{figure}[h]\label{means}
\begin{center}
\epsfig{file=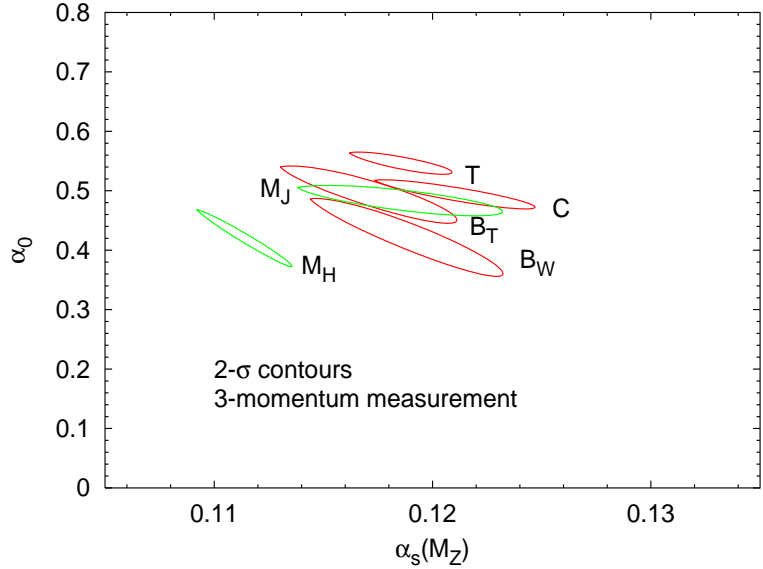,height=7.5cm,width=10cm}
   \caption{Infrared coupling from the means of jet shape variables 
(Salam \& Wicke 2000).}
\end{center}
\end{figure}
Fig.~7 verifies 
the 
(in)consistency of independent experimental determinations of the PT
and NP coupling from a variety of jet shapes, as it looks today (Salam
\& Wicke 2000). Given the relative weight of wishful thinking
substituted for rigorous proofs in formulating theoretical rules of
the game, you would agree that the hypothesis of universality (or, in
other words, the notion of the universal infrared coupling) is not
ruled out, to say the least.
(NB: two of the displayed jet shapes, namely $M_H$ and $B_W$, include
jet selection, are therefore less inclusive and may have a reason to
misbehave.)

\begin{itemize}
\item[Homework:] Divide that $\alpha_0$ by $\pi$ ($\pi=3$ would
be good enough an approximation) and please compare with the Gribov
critical coupling \eqref{QCDcrit}. 

Shall we hear the bell ringing?
\end{itemize}

\begin{acknowledgements}
I am grateful to Gavin Salam for useful remarks. 
I want to congratulate Ian Butterworth, John Ellis and Erwin
Gabathuler for the success of the Discussion Meeting they took a
burden to have organized.  I must seek their and the reader's
forgiveness for a bad quality writeup.  (Ruled by the celebrated
Reference Theorem which states that {\em The quality of someone's
paper is proportional to the number of references to your papers},
while {\em The quality of your own paper is inversely proportional to
that}.)
 
\end{acknowledgements}

%\newpage

 \end{document}